\documentstyle[preprint,eqsecnum,aps]{revtex}
\tightenlines
\input epsf

\begin{document}
\draft

\title{The breaking of the flavour permutational symmetry: Mass
   textures and the CKM matrix}

\author{A. Mondrag\'on
   and E. Rodr\'{\i}guez-J\'auregui}
\address{
  Instituto de F\'{\i}sica, UNAM, Apdo. Postal 20-364, 01000 M\'exico,
  D.F. M\'exico.}

\date{\today}
\maketitle
\begin{abstract}
 Different ans\"atze for the breaking of the flavour permutational symmetry according to $ S_{L}(3)\otimes{S_{R}(3)}\ \supset\ {S_{L}(2)\otimes S_{R}(2)}$ give different Hermitian mass matrices of the same modified Fritzsch type, which differ in the symmetry breaking pattern. In this work we obtain a clear and precise indication  on the preferred symmetry breaking scheme from a fit of the predicted 
$|{\bf V}^{th}|$ to the  experimentally determined absolute values of the elements of the $CKM$ matrix. The preferred scheme leads to simple mass textures and allows us to compute the $CKM$ mixing matrix, the Jarlskog invariant $J$, and the three inner angles of the unitarity triangle in terms of four quark mass ratios and only one free parameter: the $CP$ violating phase $\Phi$. Excellent agreement with the experimentally determined absolute values of the entries in the $CKM$ matrix is obtained for $\Phi = 90^{\circ}$. The corresponding computed values of the Jarlskog invariant and the inner angles are 
$J = 3.00\times 10^{-5}$, $\alpha= 84^{\circ}$ , $\beta= 24^{\circ}$ and $\gamma =72^{\circ}$ in very good agreement with current data on $CP$ violation in the neutral kaon-antikaon system and oscillations in the $B^{\circ}_{s}$-$\bar{B}^{\circ}_{s}$ system.
\end{abstract}

\narrowtext
\section{Introduction}

Recent interest in flavour or horizontal symmetry building (mass
textures) has been spurred mainly by the large top mass and hence, the strong hierarchy in the quark masses [1-8]. A permutational flavour symmetry has been advocated by many authors in order to constrain the fermion mass matrices and mixing parameters [9-14]. Recently, various symmetry breaking schemes have been proposed based on the discrete non-Abelian group $S_{L}(3)\otimes S_{R}(3)$, which is broken according to 
$S_{L}(3)\otimes S_{R}(3)\supset S_{L}(2)\otimes S_{R}(2)\supset S_{diag}(2)$. The group $S(3)$ treats three objects symmetrically while the hierarchical nature of the Yukawa matrices is a consequence of the representation structure, ${\bf 1 \oplus 2}$, of $S(3)$ which treats the generations differently. Different ans\"atze for the breaking of the permutational symmetry give rise to different Hermitian mass matrices ${\bf M}_{q}$ of the same modified Fritzsch type which, in a symmetry adapted basis, differ in the numerical value of the ratio $Z^{1/2} = {M_{23}\over M_{22}}$. In the absence of a physical motivated argument to fix the value of $Z^{1/2}$, different values for $Z^{1/2}$ have been proposed by various authors[1-3,6,7,18-22].

In this paper, different symmetry breaking schemes are classified in terms of the irreducible representations of an auxiliary $\tilde{S}(2)$ group. Then, diagonalizing the mass matrices, we obtain exact explicit expressions for the elements of the mixing matrix, ${\bf V}_{CKM}$, the Jarlskog invariant $J$, and the inner angles of the unitarity triangle in terms of the quark mass ratios, the simmetry mixing parameter, and one $CP$ violating phase. 
A $\chi ^{2}$ fit of the theoretical expresions to the experimentally determined absolute values of the elements of the ${\bf V}_{CKM}^{exp}$ mixing matrix gives a clear and precise indication on the preferred pattern for the breaking of the $S_{L}(3)\otimes S_{R}(3)$ symmetry. Simple, explicit expressions for the corresponding best mass textures are obtained from the best value of the mixing parameter $Z^{1/2}$. In this way we obtain an explicit parametrization of the Cabbibo-Kobayashi-Maskawa matrix in terms of the four quark mass ratios $m_{u}/m_{t},m_{c}/m_{t}, m_{d}/m_{b}, m_{s}/m_{b}$ and one $CP$ violating phase in good agreement with the experimental information about quark mixings and $CP$ violation in the $K^{\circ} - \bar{K}^{\circ}$ system and the most recent data on oscillations of the $B^{\circ}_{s}$-$\bar{B}^{\circ}_{s}$ system.

The plan of this paper is as follows: In section \ref{h2}  we review some
previous work on the breaking of the permutational flavour symmetry. A
brief group theoretical analysis of the modified Fritzsch texture is
made in section \ref{h3}. The next section is devoted to the derivation of
explicit expressions for the elements of the $CKM$ mixing matrix and the Jarlskog invariant $J$ in terms of the quark mass ratios and the mixing parameter $Z^{1/2}$. In section \ref{h5}, we extract the best value of $Z^{1/2}$ from a $\chi ^{2}$ fit of our theoretical expressions to the experimentally determined
absolute values of the entries in $|V_{CKM}^{exp}|$. The interpretation of the best value of $Z^{1/2}$ in terms of the analysis
made in sections \ref{h2} and \ref{h3} and the derivation of the corresponding best mass textures is made in section \ref{h6}. The resulting parametrization of the $CKM$ matrix in terms of four mass ratios and one $CP$ violating
phase is compared with the relevant experimental information in section 
\ref{h7}. Our paper ends with a summary of results and some
conclusions.

\section{Flavour permutational symmetry}\label{h2}

In this section, we review some previous work on the breaking of the
permutational flavour symmetry.

In the Standard Model, analogous fermions in different generations,
say ${\it u,c}$ and ${\it t}$ or ${\it d,s}$ and ${\it b}$, have
completely identical couplings to all gauge bosons of the strong, weak
and electromagnetic interactions.  Prior to the introduction of the
Higgs boson and mass terms, the Lagrangian is chiral and invariant
with respect to any permutation of the left and right quark fields.
The introduction of a Higgs boson and the Yukawa couplings give mass
to the quarks and leptons when the gauge symmetry is spontaneously
broken. The quark mass term in the Lagrangian, obtained by taking the
vacuum expectation value of the Higgs field in the quark Higgs
coupling, gives rise to quark mass matrices ${\bf M_d}$ and ${\bf
  M_u}$,

\begin{equation}\label{2}
{\cal L}_{Y} ={\bf \bar{q}}_{d,L}{\bf M}_{d}{\bf q}_{d,R}+
{\bf\bar{q}}_{u,L}{\bf M}_{u}{\bf q}_{u,R}+h.c.
\end{equation}

In this expression, ${\bf q}_{d,L,R}(x)$ and ${\bf q}_{u,L,R}(x)$ denote the
left and right quark $d$- and $u$-fields in the current or weak 
basis, ${\bf q}_{q}(x)$ is a column matrix, its components ${\bf q}_{q,k}(x)$ are the quark Dirac fields, $k$ is the flavour index. In this basis, the charged hadronic currents are

\begin{equation}
\label{4}
J_{\mu}\sim \bar{q}_{u,L}\gamma _{\mu}q_{d,L},
\end{equation}

\noindent
where

\begin{equation}
\label{5}
{\bf q}_{u,W}= \pmatrix{
u_1(x) \cr
u_2(x) \cr
u_3(x) \cr
}_{W},\hspace{1.5cm} {\bf q}_{d,W} = \pmatrix{
d_1(x) \cr
d_2(x) \cr
d_3(x) \cr
}_{W},
\end{equation}
and the subindex $W$ means weak basis.

As is evident from (\ref{4}), the charged hadronic currents are
not changed if both, the $d-$type and the $u-$type fields are
transformed with the same unitary matrix.

A number of authors [9 - 15] have pointed out that realistic quark mass
matrices result from the flavour permutational symmetry
$S_{L}(3)\otimes S_{R}(3)$ and its spontaneous or explicit breaking.
The group $S(3)$ treats three objects symmetrically, while the
hierarchical nature of the mass matrices is a consequence of the
representation structure $\bf{1\oplus2}$ of $S(3)$, which treats the
generations differently. Under exact $S_{L}(3)\otimes S_{R}(3)$
symmetry, the mass spectrum, for either up or down quark sectors,
consists of one massive particle (top and bottom quarks) in a singlet irreducible representation and a pair of massless particles in a doublet irreducible representation. In the weak basis, the mass matrix with the exact
 $S_{L}(3)\otimes S_{R}(3)$ symmetry reads
\begin{equation}\label{6}
{\bf M'}_{3q,W}= {m_{3q}\over3}\pmatrix{
1 & 1 & 1 \cr
1 & 1 & 1 \cr
1 & 1 & 1 \cr
} _{W},
\end{equation}
where $m_{3q}$ denotes the mass of the third family quark, ${\it t}$ or
${\it b}$. 

To generate masses for the second family, one has to break the
permutational symmetry $S_{L}(3)\otimes S_{R}(3)$ down to
$S_{L}(2)\otimes S_{R}(2)$. This may be done by adding to
${\bf \bar{q}}_{L}({\bf M'}_{3q,W}) {\bf q}_{R}$ a term ${\bf
  \bar{q}}_{L}({\bf M'}_{2q,W}){\bf q}_{R}$
which is invariant under $S_{L}(2)\otimes S_R(2)$ but breaks
$S_{L}(3)\otimes S_R(3)$. The most general form of a matrix ${\bf M'}_{2q,W}$ which is invariant under the permutations of the first two rows or two columns is

\begin{eqnarray}\label{8}
{\bf {M'}}_{2q,W}={m_{3q}}\pmatrix{
\alpha ' & \alpha '& \beta '\cr
\alpha '& \alpha '& \beta '\cr
\beta '& \beta '& \gamma \cr
}_{W}.
\end{eqnarray}

Without loss of generality, this matrix may be decomposed in the sum
of a $S_L(3)\otimes S_R(3)$ invariant term plus a traceless matrix ${\bf M}_{2q,W}$ invariant under $S_L(2)\otimes S_R(2)$,
\FL
\begin{eqnarray}\label{10}
{\bf M'}_{2q,W} =\frac{m_{3q}}{3}\bigg\{ (2\alpha ' + \gamma)&&\pmatrix{
1 & 1 & 1 \cr
1 & 1 & 1 \cr
1 & 1 & 1 \cr
}_{W}\cr
&+& \pmatrix{
\alpha ' - \gamma & \alpha ' - \gamma & 3\beta ' - 2\alpha ' -\gamma
\cr
\alpha ' -\gamma & \alpha ' -\gamma & 3\beta ' -2\alpha ' -\gamma \cr
3\beta ' - 2\alpha ' - \gamma & 3\beta ' -2 \alpha ' -\gamma &
-2(\alpha ' - \gamma)\cr
}_{W}\bigg\}.
\end{eqnarray}
The first term in the right hand side of (\ref{10}) is added to the term
${\bf M'}_{3q,W}$.
\begin{equation}\label{12}
{\bf M}_{3q,W} =\frac{m_{3q}}{3}(1- \Delta _q)\pmatrix{
1 & 1 & 1 \cr
1 & 1 & 1 \cr
1 & 1 & 1 \cr
}_{W}.
\end{equation}
where $\Delta _q$ stands for the factor $-(2\alpha ' +\gamma)$.

The second term in the right hand side of (\ref{10}) gives the most general form of the traceless matrix ${\bf M}_{2q,W}$ that breaks $S_L(3) \otimes S_R(3)$
down to $S_L(2)\otimes S_R(2)$ and gives mass to the second family,

\begin{equation}\label{14}
{\bf M}_{2q,W} = \frac{m_{3q}}{3}\pmatrix{
\alpha & \alpha & \beta \cr
\alpha & \alpha & \beta \cr
\beta & \beta & -2\alpha \cr
}_{W},
\end{equation}
in this expression we have simplified the notation by calling $\alpha
$ and $\beta $ in (\ref{14}), the terms $(\alpha ' - \gamma)$ and $(3\beta '
-2\alpha ' - \gamma)$ in (\ref{10}).

From expression (\ref{14}) it is evident that
${\bf M}_{2q,W}$ is a linear combination of two linearly independent numerical matrices, ${\bf M}^{A}_{2q}$  and ${\bf M}^{S}_{2q}$,

\begin{equation}\label{15}
{\bf M}_{2q,W}=\frac{m_{3q}}{3}\left( \sqrt{8}\alpha {{\bf M}_{2q,W}}^{A}+
2\beta {{\bf M}_{2q,W}}^{S}\right)
\end{equation}
where

\begin{equation}\label{16}
\begin{array}{c}
{\bf M}^{A}_{2q,W}= \frac{1}{\sqrt 8}\pmatrix {
1 & 1 & 0 \cr
1 & 1 & 0 \cr
0 & 0 & -2 \cr
}_{W}\quad\quad and \quad\quad  {\bf M}^{S}_{2q,W}=
\frac{1}{2}\pmatrix{
0 & 0 & 1 \cr
0 & 0 & 1 \cr
1 & 1 & 0 \cr
}_{W}.
\end{array}
\end{equation}
Later on, this property will be used to characterize the symmetry breaking pattern.

We may now turn our attention to the question of breaking the
$S_L(2)\otimes S_R(2)$ symmetry.
In order to give mass to the first family, we add another term
${\bf M}_{q1}$ to the mass matrix. It will be assumed that
${\bf M}_{q1}$ transforms as the mixed symmetry term of the
doublet complex tensorial representation of the $S(3)_{d}$ diagonal
subgroup of $S_{L}(3)\otimes S_{R}(3)$. Putting the first family in a
complex representation will allow us to have a CP violating phase in
the mixing matrix. Then, in the weak basis, ${\bf M}_{q1}$ is given by

\begin{equation}\label{18}
\begin{array}{c}
{\bf M}_{q1,W}={m_{3q}\over {\sqrt 3}}\pmatrix{
A_{1} & iA_{2} & -A_{1}-iA_{2} \cr
-iA_{2} & -A_{1} & A_{1}+iA_{2} \cr
-A_{1}+iA_{2} & A_1-iA_{2} & 0 \cr
}_{W} .
\end{array}
\end{equation}

Finally, adding the three mass terms, (\ref{12}), (\ref{14}) and
(\ref{18}), we get the mass matrix ${\bf M}_q$ in the weak basis.

\section{Modified Fritzsch texture}\label{h3}

To make explicit the assignment of particles to irreducible
representations of $S_{L}(3)\otimes S_{R}(3)$, it will be convenient
to make a change of basis from the weak basis to a symmetry adapted or
hierarchical basis. In this basis, the quark fields are

\begin{equation}
\label{20}
q_{1q,H}(x) = \frac{1}{\sqrt{2}} (q_{1q,W}(x) - q_{2q,W}(x)),
\end{equation}

\begin{equation}
\label{22}
{q_{2q,H}(x) = \frac{1}{\sqrt{6}} (q_{1q,W}(x) + q_{2q,W}(x) -
    2q_{3q,W}(x) )},
\end{equation}

\begin{equation}
\label{24}
{q_{3q,H}(x) = \frac{1}{\sqrt{3}} (q_{1q,W}(x) + q_{2q,W}(x) + q_{3q,W}(x)
  )},
\end{equation}

\noindent
the subindex $H$ denotes the hierarchical basis. In the hierarchical basis the third family quarks, $t$ or $b$, are assigned to the invariant singlet irreducible representation $q_{3q,H}(x)$, the other two families are assigned to $q_{2q,H}(x)$ and 
$q_{1q,H}(x)$, the two components of the doublet irreducible representation of 
$S_{diag}(3)$.

The mass matrix ${\bf M}_{q,H}$ in the symmetry
adapted basis is related to the mass matrix in the weak basis by the
unitary transformation

\begin{equation}
\label{26}
{\bf  M_{q,H}} = {\bf U}^{\dagger}{\bf M_{q,W}}{\bf U},
\end{equation}
where
\begin{equation}
\label{28}
{\bf U}={1\over \sqrt {6}}\pmatrix{
\sqrt {3} & 1 & \sqrt {2} \cr
-\sqrt {3} & 1 & \sqrt {2} \cr
0 & -2 & \sqrt {2} \cr
}.
\end{equation}

Then, in this basis, ${\bf {M}}_{q}$ takes the form
\begin{equation}
\begin{array}{rcl}
{\bf {M}}_{qH}&=&{m_{3q}}\left[ \pmatrix{
0 & {A_{q}}e^{-i\phi_{q}} & 0 \cr
{A_{q}}e^{i\phi_{q}} & 0 & 0 \cr
0 & 0 & 0 \cr
}_{H}+\pmatrix{
0 & 0 & 0 \cr
0 & -\triangle_{q}+\delta_{q} & B_{q} \cr
0 & B_{q} & \triangle_{q}-\delta_{q} \cr
}_{H}\right]\cr\cr &+&m_{3q}\pmatrix{
0 & 0 & 0 \cr
0 & 0 & 0 \cr
0 & 0 & 1-\triangle_{q} \cr
}_{H}=m_{3q}\pmatrix{
0 & A_{q}e^{-i\phi_{q}} & 0 \cr
A_{q}e^{i\phi_{q}} & -\triangle_{q}+\delta_{q} & B_{q} \cr
0 & B_{q} & 1-\delta_{q} \cr
}_{H}
\end{array}\label{30}
\end{equation}
where
\begin{equation}
\begin{array}{rcl}\label{32}
\delta_{q}=\triangle_{q}-\frac{2}{9}(\alpha+2\beta)\quad\quad and \quad\quad
B_{q}=\frac{2}{9}(\sqrt{8}\alpha - \frac{1}{\sqrt 8}\beta).
\end{array}
\end{equation}

From the strong hierarchy in the masses of the quark families, 
$m_{3q}>> m_{2q}> m_{1q}$, we expect $1-\delta_{q}$ to be very close to unity.

The entries in the mass matrix may be readily expressed in terms of the mass eigenvalues $(m_{1q}, -m_{2q}, m_{3q})$ and the small parameter $\delta_{q}$.
 Computing the invariants of $M_{q}$, $tr M_{q}$, $tr {M_{q}}^{2}$ and $det M_{q}$, we get 
\begin{eqnarray}\label{34}
A^{2}_{q}={\tilde m_{1q}}{\tilde m_{2q}}(1-\delta _{q})^{-1}\qquad ,\qquad
\triangle _{q}= {\tilde m_{2q}}-{\tilde m_{1q}}\\ \cr
B^{2}_{q}=\delta_{q}((1-\tilde m_{1q}+\tilde m_{2q}-\delta _{q})-
\tilde m_{1q}{\tilde m_{2q}}(1-\delta _{q})^{-1})
\end{eqnarray}
where 
${\tilde m_{1q}}={m_{1q}/m_{3q}}$ and 
${\tilde m_{2q}}={m_{2q}/m_{3q}}$.

If each possible symmetry breaking pattern is now characterized by the ratio
\begin{equation}
{Z_{q}}^{1/2}={B_{q}/(-\triangle _{q}+\delta_{q})},
\label{36}
\end{equation}
the small parameter $\delta _{q}$ is obtained as the solution of the cubic equation

\begin{equation}
\delta_{q}\left[ (1+\tilde m_{2q}- \tilde m_{1q}- \delta_{q})(1-\delta_{q})-
{\tilde m_{1q}}{\tilde m_{2q}}\right] - Z_{q}(1-\delta_{q})
(-{\tilde m_{2q}}+{\tilde m_{1q}}+\delta_{q})^{2}=0
\label{38}
\end{equation}
which vanishes when $Z_{q}$ vanishes. 

Equation (\ref{38}) may be written as

\begin{equation}\label{40}
\begin{array}{c}
\delta^{3}_{q} -
\frac{1}{(Z_{q}+1)}\left[Z_{q}\left(2(\tilde{m}_{2q}-\tilde{m}_{1q})+1\right)
  + (\tilde{m}_{2q}-\tilde{m}_{1q})+2\right]\delta^{2}_{q} ~~ + \\
\frac{1}{Z_{q}+1}\left[Z_{q}\left(\tilde{m}_{2q}-\tilde{m}_{1q}\right)\left(\tilde{m}_{2q}-\tilde{m}_{1q}+2\right)+\left(1-\tilde{m}_{1q}\right)\left(1+\tilde{m}_{2q}\right)\right]\delta_{q}
- \frac{Z_{q}(\tilde{m}_{2q}-\tilde{m}_{1q})^{2}}{Z_{q} + 1} = 0.
\end{array}
\end{equation}

The last term in the left hand side of (\ref{40}) is equal to the
product of the three roots of (\ref{38}). Therefore, the root of
(\ref{38}) which vanishes when $Z_{q}$ vanishes may be written as

\begin{equation}\label{42}
\delta _{q}(Z_{q}) = \frac{Z_{q}}{Z_{q}+1}\frac{
(\tilde{m}_{2q}-\tilde{m}_{1q})^{2}}{W(Z_{q})},
\end{equation}
where $W(Z_{q})$ is the product of the two roots of (\ref{40}) or
(\ref{38}) which do not vanish when $Z_{q}$ vanishes.

The product $W(Z_{q})$ is given by 

\begin{equation}\label{44}
\begin{array}{c}
W(Z_{q}) =
\left\{\left[2q^{2}+p^{3}+2q\sqrt{q^{2}+p^{3}}\right]^{1/2}+
\left[2q^{2}+p^{3}-2q\sqrt{q^{2}+p^{3}}\right]^{1/3}\right\} + \\
\frac{1}{3}\frac{1}{Z_{q}+1}\left[Z_{q}\left(2(\tilde{m}_{2q}-\tilde{m}_{1q})+
1\right)+(\tilde{m}_{2q}-\tilde{m}_{1q})+2\right]
\left\{\left[q+\sqrt{q^{2}+p^{3}}\right]^{1/3}+\left[q-\sqrt{q^{2}+p^{3}}
\right]^{1/3}\right\}\\ -|p| +
  \frac{1}{9}\frac{1}{(Z_{q}+1)^{2}}\left[Z_{q}\left(2
(\tilde{m}_{2q}-\tilde{m}_{1q})+1\right)+(\tilde{m}_{2q}-\tilde{m}_{1q})+
2\right]^{2},
\end{array}
\end{equation}
where
 
\begin{equation}\label{46}
\begin{array}{c}
2q = -\frac{2}{27}\frac{1}{(Z_{q}+1)^{3}}\left[Z_{q}\left(2(\tilde{m}_{2q} -
  \tilde{m}_{1q}) +1\right) + (\tilde{m}_{2q}-\tilde{m}_{1q}) +
  2\right]^{3}\\
+\frac{1}{3}\frac{1}{(Z_{q}+1)^{2}}
\left[Z_{q}\left(2(\tilde{m}_{2q}-\tilde{m}_{1q})+1\right)+(\tilde{m}_{2q}-
\tilde{m}_{1q}) + 2\right]\times \\
  \left\{Z_{q}(\tilde{m}_{2q}-\tilde{m}_{1q})\left((\tilde{m}_{2q}-\tilde{m}_{1q})+2\right)+(1-\tilde{m}_{1q})(1+\tilde{m}_{2q})\right\}\\ - \frac{Z_{q}}{Z_{q}+1}(\tilde{m}_{2q}-\tilde{m}_{1q})^{2},
\end{array}
\end{equation}
and
\begin{equation}\label{48}
\begin{array}{c}
3p = -\frac{1}{3}\frac{1}{(Z_{q}+1)^{2}}
\left[Z_{q}(2(\tilde{m}_{2q}-\tilde{m}_{1q})+1)+
(\tilde{m}_{2q}-\tilde{m}_{1q}) + 2\right]^{2}\\
+\frac{1}{(Z_{q}+1)}\left[Z_{q}(\tilde{m}_{2q}-\tilde{m}_{1q})(\tilde{m}_{2q}-
\tilde{m}_{1q}+2)+(1-\tilde{m}_{1q})(1+\tilde{m}_{2q})\right].
\end{array}
\end{equation}

Then, the vanishing of $Z_q$ implies that $\delta_q(Z_{q})$ vanishes and
so does $B_q$, or equivalently, there is no mixing of
singlet and doublet irreducible representations of $S_{L}(3)\otimes
S_{R}(3)$ and the heaviest quark in each sector, ${\it t}$ or ${\it b}$, is in a pure singlet representation.

\begin{figure}[htbp]
\begin{center}
    \input{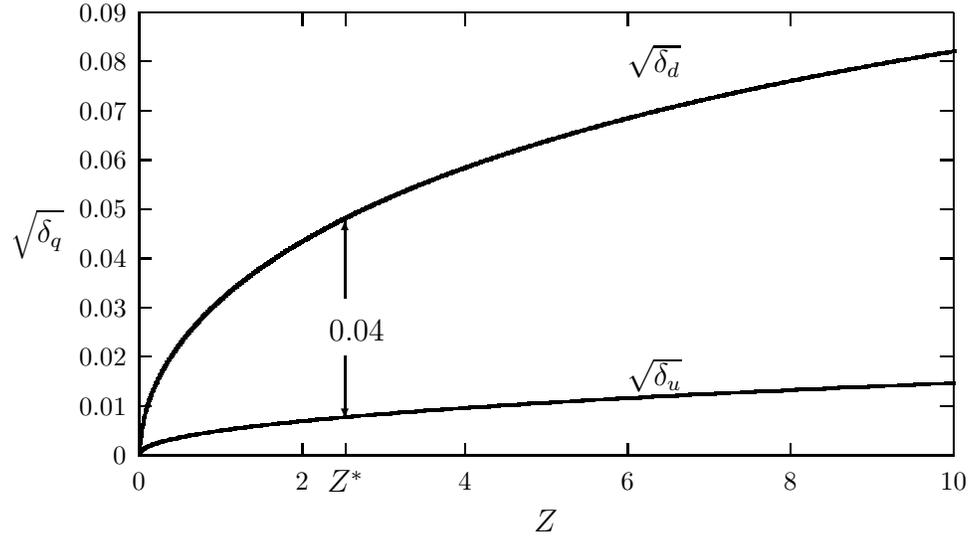}
\caption{The square root of the parameters $\delta_{u}$, $\delta_{d}$
      is shown as function of the ratio $Z_{q}$. The value $Z\approx
      5/2$ which satisfies the constraining condition (\ref{120}) may
      be read from the figure.}
    \label{Figure 1}
\end{center}
\end{figure}

In fig.~\ref{Figure 1}, ${\delta^{1/2}_{q}}$ is shown as function of $Z_{q}$. It may be seen that, as $Z_{q}$ increases, $\sqrt {\delta_{q}(Z_{q})}$
increases with decreasing curvature. For very large values of $Z_{q}$,
$\sqrt {\delta_{q}(Z_{q})}~$ goes to the asymptotic limit
$\tilde{m}_{2q}-\tilde{m}_{1q}$, 

\begin{equation}\label{50}
\lim_{z_{q} \rightarrow \infty} \delta^{1/2}_{q}(Z_{q}) = \tilde{m}_{2q}-\tilde{m}_{1q},
\end{equation}
Hence, $\delta_{q}(Z_{q})$ is a small parameter 

\begin{equation}\label{52} 
 \delta_{q}(Z_{q})<<1,  
\end{equation}
for all values of $Z_{q}$. For large values of $Z_{q}$, say $Z_{q}\geq
20$, $\delta_{q}(Z_{q})$ is not sensitive to small changes in $Z_{q}$.

From eqs. (\ref{38}) or (\ref{40}) we
derive an approximate solution for $\delta_{q}(Z_{q})$ valid for small
values of $Z_{q}$ ($Z_{q}\leq 10$). Computing in the leading order of
magnitude we obtain
\begin{eqnarray}\label{54}
 \delta_{q}\left( Z_{q} \right)\approx {Z_{q}
\left(   \tilde{m}_{2q}-\tilde{m}_{1q} \right)^{2}
\over \left(1-\tilde{m}_{1q} \right)\left( 1  +
\tilde{m}_{2q} \right)+2Z_{q}\left(   \tilde{m}_{2q}-\tilde{m}_{1q} 
\right)(1+{1\over 2}(\tilde{m}_{2q}-\tilde{m}_{1q}))}~.
\end{eqnarray}

\subsection{Symmetry breaking pattern}\label{}

In the symmetry adapted basis, ${\bf M}_{3q,H}$ is a singlet tensorial irreducible representation of   $S_{L}(3)\otimes S_{R}(3)$,

\begin{equation}\label{56}
{\bf M}_{3q,H} = m_{3q}(1- \Delta _q)\pmatrix{
0 & 0 & 0 \cr
0 & 0 & 0 \cr
0 & 0 & 1 \cr
}_{H}.
\end{equation}

\noindent 
In this same basis, the term ${\bf M}_{2q,H}$ which breaks 
$S_{L}(3)\otimes S_{R}(3)$ down to $S_{L}(2)\otimes S_{R}(2)$ is given by

\begin{equation}\label{58}
{\bf M}_{2q,H} = m_{3q}(-\Delta_{q}+\delta_{q}({Z_{q}}^{1/2}))\pmatrix{
0 & 0 & 0 \cr
0 & 1 & {Z_{q}}^{1/2}\cr
0 & {Z_{q}}^{1/2} & -1 \cr
}_{H}.
\end{equation}

The symmetry breaking pattern is characterized by the parameter ${Z_{q}}^{1/2}$
which is a measure of the mixing of singlet and doublet irreducible representations of $S_{L}(3)\otimes S_{R}(3)$. The decomposition of ${\bf M}_{2q,W}$ in a linear combination of two numerical matrices, given in eqs. (\ref{15}) and 
(\ref{16}), takes now the form

\begin{equation}\label{60}
{\bf M}_{2q,H} = m_{3q} ( -\triangle_q + \delta_q  ) \left[ 3\sqrt{2}N_{Aq} M^{A}_H + \frac{3}{2}N_{Sq} M^{S}_H \right]
\end{equation}
where the matrices 

\begin{equation}\label{62}
{\bf M}^{A}_{2,H}  = \frac{1}{3\sqrt 2}\pmatrix{ 
0 & 0 & 0 \cr
0 & 1 & -\sqrt{8} \cr
0 & -\sqrt{8} & -1 \cr
}_{H}, 
\end{equation}

\begin{equation}\label{64}
{\bf M}^{S}_{2,H}  = \frac{2}{3}\pmatrix{
0 & 0 & 0 \cr
0 & 1 & \frac{1}{\sqrt{8}} \cr
0 & \frac{1}{\sqrt{8}} & -1 \cr
}_{H},
\end{equation}
are of the same form as ${\bf M}_{2q,H}$ with mixing parameters $-\sqrt{8}$ and 
$\frac{1}{\sqrt 8}$respectively. The coefficients $N_A $ and $N_S $ are given by

\begin{equation}\label{66}
N_A = \frac{2\sqrt 2}{9}\left( \frac{1}{\sqrt 8} -Z^{1/2}_q \right)
\end{equation}
and 
\begin{equation}\label{68}
N_S = \frac{2\sqrt 2}{9}\left( \sqrt 8 +Z^{1/2}_q \right).
\end{equation}

From eqs. (\ref{58})-(\ref{64}), it is evident that there is a corresponding decomposition of the mixing parameter $Z^{1/2}_{q}$

\begin{equation}\label{70}
Z^{1/2}_{q} = N_{Aq}Z^{1/2}_{A} + N_{Sq}Z^{(q)}_{S},
\end{equation}
with
\begin{equation}\label{72}
1=N_{Aq} + N_{Sq}
\end{equation}
where $Z^{1/2}_A = - \sqrt{8}$ is the mixing parameter in the matrix 
${\bf M}^A_{2,H}$, and $Z^{1/2}_{S} = \frac{1}{\sqrt{8}}$ is the mixing parameter in ${\bf M}^S_{2,H}$. In this way, a unique linear combination of $Z^{1/2}_A$ and 
$Z^{1/2}_S$  is associated to the symmetry breaking pattern characterized by 
$Z^{1/2}_q$.

We notice that the symmetry breaking term in the Yukawa Lagrangian 
$\bar{\bf q}_L {\bf M}_{2q}{\bf q}_R $ depends only on two fields. Acording to eqs. (\ref{60}), (\ref{62}) and (\ref{64}), the term
$\bar{\bf q}_L {\bf M}_{2q,H}{\bf q}_R $ splits into the sum of a term proportional to $\bar{\bf q}_L {\bf M}^{A}_{2}{\bf q}_R $, which changes sign under the exchange of those two fields, and a term proportional to 
$\bar{\bf q}_L {\bf M}^{S}_{2}{\bf q}_R $, which remains invariant under the same exchange. Therefore, the decomposition of 
${\bf M}_{2q,H}$ given in eq. (\ref{60}) is equivalent to a decomposition of the term $\bar{\bf q}_L {\bf M}_{2q}{\bf q}_R $ into its symmetric and antisymmetric parts under the exchange of those two fields. Thus, the characterization of 
${\bf M}_{2q}$ and $Z^{1/2}_q$ as a linear combination, of ${\bf M}^{A}_2$ and 
${\bf M}^{S}_2$, and $Z^{1/2}_A$ and $Z^{1/2}_S$, respectively, (given in 
~(\ref{60})-~(\ref{72})), is equivalent to a classification of the symmetry breaking pattern defined by ${\bf M}_{2q,H}$ in terms of the irreducible representations of the group $\tilde {S}(2)$ of permutations of the two fields in $\bar{\bf q}_L {\bf M}_{2q}{\bf q}_R $.

Lehmann, Newton and Wu ~\cite{3} observed that in the case of only two families (the first two generations), the term that breaks the 
$S_{L}(2)\otimes S_{R}(2)$ symmetry changes sign when permuting the two quark fields. By analogy, they extended this observation to the case of three families and postulated that the symmetry breaking term  ${\bf M}_{2q}$ should change sign under the exchange of the two fields in 
$\bar{\bf q}_L {\bf M}_{2q}{\bf q}_R $. This assumption amounts to choosing a fixed value for the mixing parameter $Z^{1/2}_q$ equal to $-\sqrt 8$. In this paper, the pair of numbers $(N_A, N_S)$ enters as a convenient mathematical label of the symmetry breaking pattern without introducing any assumption about the actual pattern of $S_{L}(3)\otimes S_{R}(3)$ symmetry breaking realized in nature.

\section{The CKM mixing matrix}\label{h4}

The Hermitian mass matrix ${\bf M}_{q}$ may be written in terms of a
real symmetric matrix ${\bf \bar{M}}_{q}$ and a diagonal matrix of
phases ${\bf P}_{q}$ as follows
\begin{eqnarray}\label{74}
{\bf M}_{q}={{\bf P}_{q}}{\bf \bar {M}}_{q}{{\bf P}_{q}}^{\dagger}.
\end{eqnarray}

The real symmetric matrix ${\bf \bar {M}}_{q}$ may be brought to
diagonal form by means of an orthogonal transformation

\begin{eqnarray}\label{76}
{\bf \bar {M}}_{q}={{\bf O}_{q}}{\bf M}_{q~ diag}{{\bf O}_{q}}^{T},
\end{eqnarray}
where

\begin{eqnarray}\label{78}
{\bf M}_{q~ diag}=m_{3q}~diag~[\tilde{m}_{1q},-\tilde{m}_{2q},1],
\end{eqnarray}
with subscripts 1,2,3 referring to $u,c,t$ in the $u$-type sector and
$d,s,b$ in the $d$-type sector.  After diagonalization of the mass
matrices ${\bf M}_{q}$, one obtains the CKM mixing matrix as
\begin{eqnarray}\label{80}
{\bf V}_{CKM}={{\bf O}_{u}}^{T}{\bf P}^{(u-d)}{\bf O}_{d},
\end{eqnarray}
where ${\bf P}^{(u-d)}$ is the diagonal matrix of the relative phases.  

In the hierarchical basis, where ${\bf M}_{q}$ is given by eqs.(\ref{30})
-(\ref{34}), ${\bf P}^{(u-d)}$ is
\begin{eqnarray}\label{82}
{\bf P}^{(u-d)}=diag~[1,e^{i\Phi},e^{i\Phi}],
\end{eqnarray}
where

\begin{eqnarray}\label{84}
\Phi = \phi_u - \phi_d, 
\end{eqnarray}
and the orthogonal matrix ${\bf O}_{q}$ is
given by \cite{22},
\begin{equation}\label{86}
{\bf{O}}_{q}=\pmatrix{
\left(\tilde{m}_{2q}{\rm {f}}_{1}/{D}_{1} \right)^{1/2} & 
-\left(\tilde{m}_{1q}{\rm {f}}_{2}/{D}_{2} \right)^{1/2} & 
\left(\tilde{m}_{1q}\tilde{m}_{2q}{\rm {f}}_{3}/{D}_{3} 
\right)^{1/2} \cr
\left((1-\delta_{q})\tilde{m}_{1q}{\rm {f}}_{1}/{D}_{1} 
\right)^{1/2} & \left((1-\delta_{q})\tilde{m}_{2q}{\rm {f}}_{2}/{D}_{2} 
\right)^{1/2}  & \left((1-\delta_{q}){\rm {f}}_{3}/{D}_{3} 
\right)^{1/2}  \cr
-\left(\tilde{m}_{1q}{\rm {f}}_{2}{\rm {f}}_{3}/{D}_{1} 
\right)^{1/2} & -\left(\tilde{m}_{2q}{\rm {f}}_{1}
{\rm {f}}_{3}/{D}_{2} \right)^{1/2} & \left({\rm {f}}_{1}
{\rm {f}}_{2}/{D}_{3} \right)^{1/2} \cr
},
\end{equation}
where
\begin{eqnarray}\label{88}
{\rm {f}}_{1}=1-\tilde{m}_{1q}-{\delta}_{q},\quad\quad
{\rm {f}}_{2}=1+\tilde{m}_{2q}-{\delta}_{q},\quad\quad
{\rm {f}}_{3}={\delta}_{q}
\end{eqnarray}

\begin{eqnarray}\label{90}
{D}_{1}=(1-\delta_q)\left( 1-\tilde{m}_{1q} \right)\left( 
\tilde{m}_{2q}+\tilde{m}_{1q} \right),
\end{eqnarray}

\begin{eqnarray}\label{92}
{D}_{2}=(1-\delta_q)\left( 1+\tilde{m}_{2q} \right)\left( 
\tilde{m}_{2q}+\tilde{m}_{1q} \right),
\end{eqnarray}

\begin{eqnarray}\label{94}
{D}_{3}=(1-\delta_q)\left( 1+\tilde{m}_{2q} \right)\left( 
1-\tilde{m}_{1q} \right).
\end{eqnarray}

From eqs. (\ref{74}-\ref{94}), all entries in the ${\bf V}_{CKM}$
matrix may be written in terms of four mass ratios: 
$(\tilde{m}_{u},\tilde{m}_{c}, \tilde{m}_{d}, \tilde{m}_{s})$ and three free real parameters : $\delta_{u}, \delta_{d}$ and $\Phi=\phi_{u}-\phi_{d}$.
The phase $\Phi$ measures the mismatch in the $S_{L}(2)\otimes S_{R}(2)$
symmetry breaking in the $u$- and $d$-sectors. In
this picture of the ${\bf V}_{CKM}$ matrix, it is this phase, and
consequently, that mismatch, which is responsible for the 
violation of CP.

The Jarlskog invariant , $J$, may be computed directly from the commutator of the mass matrices ~\cite{26}

\begin{equation}\label{96}
J=- \frac{det \{ -i[{\bf M}_{u,H}, {\bf M}_{d,H}]\} }{F}
\end{equation}
where 

\begin{equation}\label{98}
F=(1+\tilde{m}_{c})(1-\tilde{m}_{u})(\tilde{m}_{c}+\tilde{m}_{u})
(1+\tilde{m}_{s})(1-\tilde{m}_{d})(\tilde{m}_{s}+\tilde{m}_{d}).
\end{equation}

Substitution of the expression (\ref{30}) for ${\bf M}_u$ and ${\bf M}_d$, in (\ref{96}), with $Z^{1/2}_u=Z^{1/2}_d=Z^{1/2}$, gives

\begin{eqnarray}
&J&={{Z  \sqrt{{\tilde m_{u}/ \tilde m_{c}}
\over{1-\delta_{u}}} \sqrt{{\tilde m_{d}/ \tilde m_{s}}
\over {1-\delta_{d}}}sin{\Phi}}\over{(1+\tilde m_{c})
(1-\tilde m_{u})(1+\tilde m_{u}/ \tilde m_{c})
(1+\tilde m_{s})(1-\tilde m_{d})(1+\tilde m_{d}/ \tilde m_{s})}}\cr
& \times &\bigg\{
 [(-\triangle_{u}+\delta_{u})(1-\delta_{d})-(-\triangle_{d}+\delta_{d})(1-\delta_{u})]^{2}-{\left({\tilde m_{u}\tilde m_{c}}\over {1-\delta_{u}}\right)}(-\triangle_{d}+\delta_{d})^{2}\cr & - & {\left({\tilde m_{d}\tilde m_{s}\over {1-\delta_{d}}}\right)}(-\triangle_{u}+\delta_{u})^{2}+2 \sqrt{\tilde m_{u}\tilde m_{c}\over {1-\delta_{u}}}\sqrt{\tilde m_{d}\tilde m_{s}\over {1-\delta_{d}}}(-\triangle_{u}+\delta_{u})(-\triangle_{d}+\delta_{d})cos{\Phi}\bigg\}.
\label{100}
\end{eqnarray}

Explicit expressions for $\Delta_q$ and $\delta_q$ in terms of the quark masses are given in eqs. ~(\ref{34}) and ~(\ref{42})-(\ref{48}).

In this way, an exact closed expression for $J$ in terms of the quark masses, the symmetry breaking parameter $Z$ and the $CP$ violating phase $\Phi$ is obtained. Let us recall that the non-vanishing of $J$ is a necessary and sufficient condition for the violation of $CP$ ~\cite{26}. From eq. (\ref{100}), it is apparent that $J$ vanishes when $Z$, $sin \Phi$ and $\tilde{m}_u$ or $\tilde{m}_d$ vanish. Therefore, the violation of $CP$ and the consequent non-vanishing of 
$Z$ necessarily implies a mixing of singlet and doublet representations of $S_{L}(3)\otimes S_{R}(3)$.

\section{The best value of ${Z^{1/2}_{q}}$ }\label{h5}

At this stage in our argument, a question comes naturally to mind.
Does a comparison of the theoretical mixing matrix ${\bf V}^{th}_{CKM}$
with the experimentally determined ${\bf V}^{exp}_{CKM}$ give any clue
about the actual pattern of $S_{L}(3)\otimes S_{R}(3)$ symmetry
breaking realized in nature? or phrased differently: What are the best
values for $Z_{u}$ and $Z_{d}$? What is the best value for $\Phi$? Do
these values correspond to some well defined symmetry breaking
pattern?  

As a first step in the direction of finding an answer
to these questions, we made a $\chi^{2}$ fit of the exact expressions
for the absolute value of the entries in the mixing matrix, that is
$|{V}^{th}_{CKM}|$ and the Jarlskog invariant $J^{th}$ to the
experimentally determined values of $|{V}^{exp}_{CKM}|$ and $J^{exp}$.
Since the value of the observed CKM matrix parameters $|{\bf V}^{exp}_{CKM}|$ are given at $\mu = m_t $, in the calculation we used the values of the running quark masses evaluated at $m_{t}$. These values were taken from the work of Fritzsch \cite{16}, see also Fusaoka and Koide ~\cite{15} and F. Yndur\'ain 
~\cite{27}. We kept the mass ratios 
$\tilde {m}_{c}=\frac {m_c}{m_t}$ and $\tilde {m}_{s}=\frac {m_s}{m_b}$ fixed at their central values

\begin{eqnarray}\label{102}
{\tilde{m}}_{c}=0.0048\qquad and\qquad{\tilde{m}}_{s}=0.03437
\end{eqnarray}
but, for reasons which will be apparent later, we took the values
\begin{eqnarray}\label{104}
{\tilde{m}}_{u}=0.000042\qquad and\qquad{\tilde{m}}_{d}=0.00148 ,
\end{eqnarray}
which are close to the upper and lower bounds of 
$\tilde {m}_u = \frac{m_u}{m_t}$ and $\tilde {m}_d = \frac{m_d}{m_b}$ respectively, and we looked for the best values of the three parameters 
$\delta_{u},\delta_{d}$ and $\Phi$.  We found the following results \cite{25}:

I.- Excellent fits of similar quality, $\chi^{2}\leq 0.33$, were
  obtained for a continuous family of values of the parameters
  $(\delta_{u},\delta_{d})$.

II.- In each good quality fit, the best value of $\Phi$ was
  fixed without ambiguity.

III.- The best value of $\Phi$ was nearly stable against large
changes in the values of $(\delta_{u},\delta_{d})$ which produced fits
of the same good quality.

IV.- In all good quality fits, the difference
$\sqrt{\delta_{d}}-\sqrt{\delta_{u}}$ takes the same value
\begin{eqnarray}\label{106}
\sqrt{\delta_{d}}-\sqrt{\delta_{u}}\simeq 0.040 .
\end{eqnarray}

These results may be understood if we notice that not all entries in
${{\bf V}_{CKM}}^{th}$ are equally sensitive to variations of the different
parameters. Some entries, like $V_{us}$, are very sensitive to changes
in $\Phi$ but are almost insensitive to changes in
$(\delta_{u},\delta_{d})$ while, some others, like $V_{cb}$ are almost
insensitive to changes in $\Phi$ but depend critically on the
parameters $\delta_{u}$ and $\delta_{d}$.

From eqs. (\ref{74})-(\ref{94}), we obtain
\begin{eqnarray}\label{107}
&V_{us}&=-\left( {{\tilde{m}}_{c}{\tilde{m}}_{d}\over 
{\left( 1-{\tilde{m}}_{u} \right)\left( {\tilde{m}}_{c}+
{\tilde{m}}_{u} \right)\left( 1+{\tilde{m}}_{s}\right)
\left(  {\tilde{m}}_{s}+{\tilde{m}}_{d}\right)}}\right)^{1/2} 
\left({\left(1-{\tilde{m}}_{u}-{\delta}_{u}  \right) \left(1+
{\tilde{m}}_{s}-{\delta}_{d}\right)\over {\left( 1-{\delta}_{u} 
\right) 
\left(1-{\delta}_{d} \right)}}\right)^{1/2}\cr 
&+&\bigg\{ \left(\frac {\left(1-\tilde{m}_{u}- \delta_{u} \right)
\left(1+\tilde{m}_{s}- \delta_{d} \right)}{\left( 1+\tilde{m}_{s} \right)}
\right)^{1/2}+\left(\frac{ \left(1+\tilde{m}_{c}- \delta_{u} \right)\delta_u}
{1-\delta_u}\right)^{1/2}\left(\frac{ \left(1-\tilde{m}_{d}- \delta_{d} \right)\delta_d}{(1-\delta_d)(1+\tilde{m}_s)}\right)^{1/2}\bigg\}\cr &\times&
\left( {{\tilde{m}}_{u}{\tilde{m}}_{s}
\over {\left( 1-{\tilde{m}}_{u} \right)\left( {\tilde{m}}_{c}+
{\tilde{m}}_{u} \right)\left(  {\tilde{m}}_{d}+{\tilde{m}}_{s}
\right)}} \right) ^{1/2}e^{i\Phi}. 
\end{eqnarray}
In the leading order of magnitude,
\begin{eqnarray}\label{108}
\mid V_{us} \mid \approx \mid\sqrt{{\tilde m}_{d} / {\tilde m}_{s}} 
- \sqrt{{\tilde m}_{u} / {\tilde m}_{c}} e^{i\Phi}
\mid\left( 1+ {\tilde m}_{u} / {\tilde m}_{c}+{\tilde m}_{d} / 
{\tilde m}_{s}\right)^{-1/2}.
\end{eqnarray}
Hence,
\begin{eqnarray}\label{110}
\cos \Phi ~\approx {{{\tilde m}_{d} / {\tilde m}_{s}+
{\tilde m}_{u} / {\tilde m}_{c}-\mid V_{us} \mid^{2}\left(  1+ 
{\tilde m}_{u} / {\tilde m}_{c}+{\tilde m}_{d} / {\tilde m}_{s}
\right)}\over {2 \sqrt{({\tilde m}_{d} / {\tilde m}_{s})
({\tilde m}_{u} / {\tilde m}_{c})} }}.
\end{eqnarray}
Substitution of $|{V_{us}}^{exp}|^{2}$ for $|V_{us}|^{2}$ and the
numerical value of the mass ratios, (\ref{102}) and (\ref{104}), in
(\ref{110}) gives
\begin{eqnarray}\label{112}
87^{\circ}\leq\Phi\leq 92^{\circ}
\end{eqnarray}
with a mean value 
\begin{eqnarray}\label{114}
\bar{\Phi}=89.5^{\circ}~,
\end{eqnarray}
in good agreement with the best value extracted from the preliminary
$\chi^{2}$ fits \cite{25}.

Similarly, ${V_{cb}}^{th}$ is given by
\begin{eqnarray}\label{116}
{V_{cb}}^{th}&=&- \left( \frac{\tilde {m}_u (1+\tilde{m}_c -\delta_u ) }
{ (1-\delta_u)(1+\tilde{m}_c)(\tilde{m}_c + \tilde{m}_u)} \frac {
\tilde {m}_d \tilde {m}_s \delta_d }{(1- \delta_d )(1+ \tilde {m}_s)
(1 - \tilde {m}_d) } \right)^{1/2}\cr
&+& \bigg\{ \left(\frac{\tilde{m}_c (1+\tilde{m}_c -\delta_u)}
{(\tilde{m}_c +\tilde{m}_u)(1+\tilde{m}_c)}\frac{\delta_d}{(1+\tilde{m}_s)
(1-\tilde{m}_d)}\right)^{1/2} \cr
&-& \left(\frac{\tilde{m}_c (1-\tilde{m}_u -\delta_u)\delta_u 
(1-\tilde{m}_d -\delta_d)(1+\tilde{m}_s -\delta_d)}
{(1- \delta_u )(1+\tilde{m}_c)(\tilde{m}_c +\tilde{m}_u)(1- \delta_d)
(1+\tilde{m}_s)(1-\tilde{m}_d)}
\right)^{1/2}\bigg\}e^{i\Phi}.
\end{eqnarray}
Therefore, in the leading order of magnitude, $|V_{cb}|$ is
independent of $\Phi$ and given by
\begin{eqnarray}\label{118}
\mid {V}_{cb} \mid\approx\sqrt{\delta_{d}} -\sqrt{\delta_{u}}~.
\end{eqnarray}
Hence, good agreement with $|{V_{cb}}^{exp}|\approx 0.039$ \cite{24}
requires that
\begin{eqnarray}\label{120}
\sqrt{\delta_{d}} -\sqrt{\delta_{u}}\approx 0.040,
\end{eqnarray}
at least for one pair of values $(\delta_{u},\delta_{d})$.  

Finally, let us notice that the matrix elements $V_{ub}$ and $V_{dt}$, as well as the Jarlskog invariant (see eq. (\ref{100})), are sensitive to small changes in the masses of the light quarks $\tilde {m}_u$ and $\tilde {m}_d$. For instance, 

\begin{eqnarray}\label{122}
V_{ub}&=&\left( \frac{\tilde {m}_c (1-\tilde{m}_u -\delta_u ) }{ (1-\delta_u)
(1-\tilde{m}_u)(\tilde{m}_c + \tilde{m}_u)} \frac {\tilde {m}_d \tilde {m}_s
\delta_d }{(1- \delta_d )(1+ \tilde {m}_s)(1 - \tilde {m}_d) } \right)^{1/2}
\cr
&+& \bigg\{ \left(\frac{\tilde{m}_u (1-\tilde{m}_u -\delta_u)\delta_d}
{(1-\tilde{m}_u)(\tilde{m}_c +\tilde{m}_u)(1+\tilde{m}_s)(1-\tilde{m}_d)}
\right)^{1/2} \cr
&-& \left(\frac{\tilde{m}_u (1+\tilde{m}_c -\delta_u)\delta_u 
(1-\tilde{m}_d -\delta_d)(1+\tilde{m}_s -\delta_d)}
{(1- \delta_u )(1-\tilde{m}_u)(\tilde{m}_c +\tilde{m}_u)(1- \delta_d)
(1+\tilde{m}_s)(1-\tilde{m}_d)}
\right)^{1/2}\bigg\}e^{i\Phi}  
\end{eqnarray}
computing in the leading order of magnitude, we get

\begin{equation}\label{124}
V_{ub}\approx \sqrt{\frac{\tilde{m}_u}{\tilde{m}_c}}\left( \sqrt{\delta_d} - 
\sqrt{\delta_u}\right)e^{i\Phi}.
\end{equation} 

A similar computation gives for $V_{td}$ 

\begin{equation}\label{126}
V_{td}\approx -\sqrt{\frac{\tilde{m}_d}{\tilde{m}_s}}\left( \sqrt{\delta_d} - 
\sqrt{\delta_u}\right)e^{i\Phi}.
\end{equation}

However, since the masses of the light quarks are the less well determined, and the moduli $|V^{exp}_{ub}|$ and $|V^{exp}_{td}|$ have the largest error bars, relatively large changes in the values of $\tilde {m}_u$ and  $\tilde {m}_d$ produce only very small variations in the goodness of fit of the thoretical matrix of moduli $|V^{th}|$ to the experimentally determined $|V^{exp}|$. The sensitivity of the matrix elements $|V_{ub}|$ and $|V_{td}|$ to changes in 
$\tilde {m}_u$ and  $\tilde {m}_d$ is reflected in the shape of the unitarity triangle which changes appreciably when the masses of the light quarks change within their uncertainty bounds, as may be seen from the following expressions

\begin{eqnarray}\label{128}
\alpha = arg \left(-\frac{V^{*}_{ub}V_{ud}}{V^{*}_{tb}V_{td}}\right)
\approx arg \left\{ \left(\frac{\tilde{m}_u}{\tilde{m}_c}\frac{\tilde{m}_s}
{\tilde{m}_d}\right)^{1/2}e^{-i\Phi}\right\}= \Phi,
\end{eqnarray}

\begin{eqnarray}\label{130}
\beta = arg \left(-\frac{V^{*}_{tb}V_{td}}{V^{*}_{cb}V_{cd}}\right)\approx
arctan\bigg\{ \frac{\sqrt{\frac{\tilde{m}_u}{\tilde{m}_c}}sin \Phi}{\sqrt
{\frac{\tilde{m}_d}{\tilde{m}_s}}-\sqrt{
\frac{\tilde{m}_u}{\tilde{m}_c}}cos \Phi}\bigg\},
\end{eqnarray}
and

\begin{eqnarray}\label{132}
\gamma = arg \left(-\frac{V^{*}_{cb}V_{cd}}{V^{*}_{ub}V_{ud}}\right)\approx
arctan\bigg\{ \frac{\sqrt{\frac{\tilde{m}_d}{\tilde{m}_s}}sin \Phi}{\sqrt{
\frac{\tilde{m}_u}{\tilde{m}_c}}-\sqrt{\frac{\tilde{m}_d}{\tilde{m}_s}}cos\Phi}
\bigg\},
\end{eqnarray}
$\alpha$, $\beta$ and $\gamma$ are the inner angles of the unitarity triangle. When the central values of $\tilde {m}_u=0.000018$ and $\tilde {m}_d=0.0019$ 
\cite{16} are used in the fitting procedure, the agreement of $|{\bf V}^{th}|$ with $|{\bf V}^{exp}|$ is very good, $\chi^{2}=0.33$, but we systematically obtain $\gamma^{th} > \alpha^{th}$ in stark disagreement with the most recent data on the $K^{\circ}-\bar{K}^{\circ}$ system and the most recent data on the 
$B^{\circ}_{s,d}$ oscillations ~\cite{24} and ~\cite{28}. We could not change the values of $\sqrt{\delta_d} - \sqrt{\delta_u}$ without spoiling the good overall agreement of $|{\bf V}^{th}|$ to 
$|{\bf V}^{exp}|$. Therefore, we let the masses of the light quarks vary within their uncertainty ranges. The best simultaneous $\chi^{2}$ fit of $|V^{th}|$, $J^{th}$ and 
$\alpha^{th}$, $\beta^{th}$ and $\gamma^{th}$ to the experimentally determined quantities $|V^{exp}|$, $J^{exp}$ and $\alpha^{exp}$, $\beta^{exp}$ and 
$\gamma^{exp}$ ~\cite{24}, ~\cite{28} was obtained when the value of 
$\tilde{m}_u$ is taken close to its upper bound, $\tilde{m}_u\approx 0.000042$, and the value of $\tilde{m}_d\approx 0.00148$, which is close to its lower bound. Notice that, the chosen high value of $\tilde{m}_u$ gives for the ratio 
$|V_{ub}|/|V_{cb}|$ the value
 
\begin{equation}\label{133}
\frac{|V_{ub}|}{|V_{cb}|}\approx \sqrt{\frac{\tilde{m}_u}{\tilde{m}_c}}=0.093
\end{equation}
in very good agreement with its latest world average \cite{28}.

We may now return to our discussion of the determination of the best pattern of symmetry breaking. 
As explained above, in the preliminary $\chi^{2}$ fit to the data it was found that $\sqrt{\delta_d}-\sqrt{\delta_u}\approx 0.04$, eq. ~(\ref{120}), is satisfied almost exactly even when we let the masses of the light quarks vary, not just for one pair of values $(\delta_u , \delta_d)$ but for a continuous range of values of $\delta_u$ and $\delta_d$ in which these parameter change by more than one order of magnitude.

Therefore, eq.~(\ref{120}) may be used as a constraining condition
on the possible values of $(\delta_{u},\delta_{d})$.  In this way, we
eliminate one free parameter in ${{\bf V}_{CKM}}^{th}$ without spoiling the
good quality of the fit. 
However, fixing the numerical value of this free parameter is not
enough to get a clear indication on what is the actual pattern of flavour
symmetry breaking realized in nature. This is so because
according to eqs.(\ref{38})-(\ref{48}), $\delta_{q}$ is a function of the mass
ratios $(\tilde{m}_{1q},\tilde{m}_{2q})$ and the parameter
$Z^{1/2}_{q}$ which caracterizes the pattern of $S_{L}(3)\otimes
S_{R}(3)$ symmetry breaking in the $q-$sector. Hence, a convenient way
to isolate the information about the pattern of symmetry breaking
carried by the constraining conditions (\ref{120}) from the information
on the numerical values of the quark mass ratios, is to change the
parametrization of ${\bf V}_{CKM}^{th}$ by writing $\delta_{q}$ as function
of $Z^{1/2}_{q}$ with fixed values of
($\tilde{m}_{1q},\tilde{m}_{2q}$). In this way ${\bf V}_{CKM}^{th}$ becomes
a function of the two free parameters ($Z^{1/2}_{u},Z^{1/2}_{d}$)
instead of ($\delta_{u},\delta_{d}$).

A simple approximate expression for the constraining condition
(\ref{106}), (\ref{120}) in terms of ($Z^{1/2}_{u},Z^{1/2}_{d}$), valid for 
$0\leq Z_{q}\leq 10$, is obtained from (\ref{120}), writing $\delta_{q}(Z_{q})$ in the leading order of magnitude
\begin{eqnarray}\label{134}
\sqrt{\delta_{d}}-\sqrt{\delta_{u}}\simeq&&
{{Z_{d}}^{1/2}\left( \tilde{m}_{s}-\tilde{m}_{d}\right)
\over \sqrt { \left( 1+\tilde{m}_{s}\right)
\left(  1-\tilde{m}_{d}\right)+2Z_{d} \left(
  \tilde{m}_{s}-\tilde{m}_{d} 
\right)}}\cr - &&{{Z_{u}}^{1/2}
\left( \tilde{m}_{c}-\tilde{m}_{u}\right)
\over \sqrt { \left( 1+\tilde{m}_{c}\right)
\left(  1-\tilde{m}_{u}\right)+2Z_{u} 
\left( \tilde{m}_{c}-\tilde{m}_{u} \right)}}\simeq0.040
\end{eqnarray}
When the condition (\ref{134}) is satified, to each value of
$Z_{u}^{1/2}$ corresponds one value of $Z_{d}^{1/2}$. But, since we
have only one condition to fix the value of two parameters,
$Z_{u}^{1/2}$ would still be a free parameter. Therefore, to avoid this continuous
ambiguity, we will further  assume that the up and down mass matrices
are generated following the same symmetry breaking pattern, that is,
\begin{eqnarray}\label{136}
Z_{u}^{1/2}=Z_{d}^{1/2} \equiv Z^{1/2}.
\end{eqnarray}
Then, the value of $Z$ which satisfies the constraining conditions (\ref{120}) and (\ref{134}) may be read directly from Fig. 1. We find $Z^{*}\simeq2.5$.

A more precise numerical computation of the best value of $Z$ was made
using the exact numerical solution of eq.~(\ref{38}), given in
eqs.(\ref{42}) - (\ref{48}), to compute the entries in ${\bf V}_{CKM}^{th}$
as functions of only two free parameters, $\Phi$ and $Z^{1/2}$. As
previously, we kept the mass ratios fixed at the values given in
(\ref{102}) and (\ref{104}). Then, we made a new $\chi ^{2}$ fit of the
exact expressions for the absolute values of the entries in the
theoretical expressions for $|V_{CKM}^{th}|$ and the Jarlskog
invariant $J^{th}$, to the experimentally determined values of
$|V_{CKM}^{exp}|$ and $J^{exp}$. We found the following best values
for $\Phi$ and $Z$,

\begin{equation}\label{138}
\Phi = 89.3^{\circ},
\end{equation}
and

\begin{equation}\label{140}
2.40 \leq Z^{*} \leq 2.55,
\end{equation}
corresponding to a value of $\chi ^{2} \leq 0.33.$

When the best value of the CP violating phase $\Phi=89.33^{\circ}$ is changed by one degree, the computed values of all entries in the matrix of moduli 
$|V^{th}_{CKM}|$ change in the fourth decimal place, except $|V^{th}_{us}|$ and
 $|V^{th}_{cd}|$ which change in the third decimal place by an amount which is about one fourth of the uncertainty in the experimentally determined values of 
$|V^{exp}_{us}|$ and $|V^{exp}_{cd}|$ as reported in C. Caso et al \cite{24}.
The value of $\chi^{2}$ changes from 0.33 to 0.44 which is not statistically significative. Therefore, we will adopt as the best value of $\Phi$ the simple figure
\begin{equation}\label{144}
\Phi^{*}=90^{\circ}
\end{equation}
Once the best value of $Z$ has been found, the question posed at the
beginning of this section takes a new form: What is the symmetry
breaking pattern correspondieng to $Z^{*} \simeq 2.5  ?$

An answer would be readily found if $Z^{*1/2}$ could be written as a
simple, non-trivial, linear combination of $Z^{1/2}_{A}$ and
$Z^{1/2}_{S}$, which, are equal to $-\sqrt{8}$ and $1/{\sqrt{8}}$ respectively.
From these numbers, we find that $Z^{*1/2}$ may indeed be written as 

\begin{equation}\label{148}
Z^{*1/2} = \frac{1}{2}\left[Z^{1/2}_{s} - Z^{1/2}_{A}\right] =
\frac{1}{2}\left[1/\sqrt{8} + \sqrt{8}\right],
\end{equation}
then

\begin{equation}\label{150}
Z^{*} = \frac{81}{32} = 2.53125 .
\end{equation}
The corresponding values of $\delta_{u}(Z)$ and $\delta_{d}(Z)$ are
\begin{eqnarray}\label{142}
\delta_{u}(Z^{*})=0.000056,\qquad\qquad\qquad\delta_{d}(Z^{*})=0.0023~~.
\end{eqnarray}

Let us remark again that the numerical value of $Z^{*1/2}$ was
extracted from a fit of $|V_{CKM}^{th}|$ to the experimentally
determined absolute values of the elements of the $CKM$ mixing matrix.
The identification of $Z^{*1/2}$ with the expression (\ref{148}) gives
a clear and precise indication about the preferred pattern for the
breaking of the $S_{L}(3)\otimes S_{R}(3)$ permutational flavour
symmetry by the quark mass matrices.

\section{Mass textures from the ``best'' symmetry breaking scheme}\label{h6}

Once the best value of $Z^{1/2}$ has been determined, we may turn the
argument around, and propose it as a symmetry breaking ansatz in the
form of the following assumption. The $S_{L}(3)\otimes S_{R}(3)$
flavour symmetry is broken down to $S_{L}(2)\otimes S_{R}(2)$
according to a mixed symmetry breaking pattern, which, in the
hierarchical basis, is characterized by

\begin{equation}\label{152}
Z^{*1/2} = \frac{1}{2}\left(Z^{1/2}_{S}-Z^{1/2}_{A}\right).
\end{equation}
Then, the mass matrix with the modified Fritzsch texture takes the
form

\begin{equation}\label{154}
{\bf M}^{*}_{q,H} = m_{3q}\pmatrix{
0      &
\sqrt{\frac{\tilde{m}_{1q}\tilde{m}_{2q}}{1-\delta^{*}_{q}}}e^{-i\phi_{q}}
 & 0 \cr
\sqrt{\tilde{m}_{1q}\tilde{m}_{2q}\over{1-\delta^{*}_{q}}}e^{i\phi_{q}} &
-\tilde{m}_{2q}+\tilde{m}_{1q}+\delta^{*}_{q} &
\frac{9\sqrt{2}}{8}(-\tilde{m}_{2q}+\tilde{m}_{1q}+\delta^{*}_{q}) \cr
0 & \frac{9\sqrt{2}}{8}(-\tilde{m}_{2q}+\tilde{m}_{1q}+\delta^{*}_{q})
& 1-\delta^{*}_{q} \cr
}_{H},
\end{equation}
where $\delta^{*}_{q}$ is the solution of the cubic equation

\begin{equation}\label{156}
\begin{array}{c}
113\delta^{*3}_{q}-\left[194\left(\tilde{m}_{2q}-\tilde{m}_{1q}\right)
  + 145\right]\delta^{*2}_{q} + \\
  \left[81\left(\tilde{m}_{2q}-\tilde{m}_{1q}\right)^{2}+
194\left(\tilde{m}_{2q}-\tilde{m}_{1q}\right)-
32\tilde{m}_{1q}\tilde{m}_{2q}+32\right]\delta^{*}_{q}-
81\left(\tilde{m}_{2q}-\tilde{m}_{1q}\right)^{2} = 0,
\end{array}
\end{equation}
obtained from eq. (\ref{40}) when $\sqrt{81/32}$ is
subtituted for $Z^{*1/2}$.

The mass matrix ${\bf M}_{q,H}$ was built up adding three terms, ${\bf
  M}_{1q,H}, {\bf M}_{2q}$ and ${\bf M}_{3q,H}$. The term ${\bf
  M}_{3q,H}$ is a singlet irreducible representation of
$S_{L}(3)\otimes S_{R}(3)\supset S_{diag}(3)$

\begin{equation}\label{158}
{\bf M}_{3q,H} = \left(m_{3q}-m_{2q}+m_{1q}\right)\pmatrix{
0 & 0 & 0 \cr
0 & 0 & 0 \cr
0 & 0 & 1 \cr
}_{H}.
\end{equation}

The matrix ${\bf M}_{2q,H}$ breaks $S_{L}(3)\otimes S_{R}(3)$ down to
$S_{L}(2)\otimes S_{R}(2)$, mixing the singlet and doublet
representation of $S_{diag}(3)$ in a proportion precisely determined
by the mixing parameter $Z^{*1/2}=\sqrt{\frac{81}{32}}$,

\begin{equation}\label{160}
{\bf M}^{*}_{2q,H} =
m_{3q}\left(-\tilde{m}_{2q}+\tilde{m}_{1q}+\delta^{*}_{q}\right)\pmatrix{
0 & 0 & 0 \cr
0 & 1 & \sqrt{81/32} \cr
0 & \sqrt{81/32} & -1 \cr
}_{H}.
\end{equation}
The mixing parameter $\sqrt{81/32}$ corresponds to what was called in
section \ref{h2}, $|V^{th}_{CKM}|$ a mixed symmetry breaking pattern, that is, it may be split
in the sum of a term ${\bf M}^{A}_{2q,H}$ corresponding to a purely
antisymmetric, plus a term ${\bf M}^{S}_{2q,H}$ corresponding to a purely
symmetric breaking pattern. The coefficients in each term, $N_{S} =
\frac{25}{18}$ and $N_{A} = -\frac{7}{18}$, are obtained solving the
pair of coupled equations (\ref{70}) and (\ref{72}) when $Z^{*1/2} =
\sqrt{\frac{81}{32}}.$

Hence,

\begin{equation}\label{162}
{\bf M}^{*}_{2q,H}=m_{3q}\left(-\tilde{m}_{2q}+\tilde{m}_{1q}+\delta^{*}_{q}\right)\left[\frac{-7}{18}\pmatrix{
0 & 0 & 0 \cr
0 & 1 & -\sqrt{8} \cr
0 & -\sqrt{8} & -1 \cr
}_{H} +
\frac{25}{18}
\pmatrix{
0 & 0 & 0 \cr
0 & 1 & \frac{1}{\sqrt{8}} \cr
0 & \frac{1}{\sqrt{8}} & -1 \cr
}_{H}\right].
\end{equation}
The $S_{L}(2)\otimes S_{R}(2)$ symmetry of this term and its splitting
in the sum of a purely symmetric plus a purely antisymmetric breaking
pattern terms is evident in the weak representation

\begin{equation}\label{164}
{\bf M}^{*}_{2q,W} =
m_{3q}\left(-\tilde{m}_{2q}+\tilde{m}_{1q}+\delta^{*}_{q}\right)\left(\frac{1}{24}\right)\pmatrix{
14 & 14 & -25 \cr
14 & 14 & -25 \cr
-25 & -25 & -28 \cr
}_{W}.
\end{equation}

Finally, the term ${\bf M}_{1q,H}$ breaks the $S_{L}(2)\otimes S_{R}(2)$
symmetry

\begin{equation}\label{166}
M^{*}_{1q,H}=
m_{3q}\sqrt{\frac{\tilde{m}_{1q}\tilde{m}_2q} {1-\delta^{*}_{q}}}\pmatrix{
0 & e^{-i\phi_{q}} & 0 \cr
e^{i\phi_{q}} & 0 & 0 \cr
0 & 0 & 0 \cr
}_{H}.
\end{equation}
Since $\delta^{*}_{q}$ is a function of the mass ratios
$(\tilde{m}_{1q},\tilde{m}_{2q})$ the phase $\phi_{q}$ is the only
free parameter left in the mass matrix ${\bf M}^{*}_{q}$.

\section{The mixing matrix, $V_{CKM}$, from the best symmetry breaking
  scheme}\label{h7}

We have seen that, once the symmetry breaking ansatz fixes the value
of the mixing parameter $Z^{1/2}$ at $\sqrt{81/32}$, the entries in the
mass matrices ${\bf M}_{q}$ are functions of the mass ratios
$(\tilde{m}_{1q},\tilde{m}_{2q})$ and the phase $\phi_{q}$ which is a
free parameter.

After factorizing the phases, as in eq.(\ref{74}), all entries in the
real symmetric matrices ${\bf \bar{M}}_{q}$ are functions of the mass
ratios $(\tilde{m}_{1q},\tilde{m}_{2q})$ only. Hence, the orthogonal
matrices ${\bf O}_{q}$ which bring ${\bf \bar{M}}_{q}$ to diagonal
form are also functions of $(\tilde{m}_{1q},\tilde{m}_{2q})$ only.

According to eq.(\ref{80}), ${\bf V}_{CKM}$ is given by ${\bf
  O^{T}_{u}P^{(u-d)}O_{d}}$, where ${\bf P^{(u-d)}}$ is the diagonal
matrix of the relative phases.
Therefore, once the symmetry breaking ansatz determines the value of
$Z^{*1/2} = \sqrt{\frac{81}{32}}$, the theoretical expressions for the
entries in the mixing matrix, ${\bf V}_{CKM}^{th}$, are written in terms of
the four mass ratios
$(\tilde{m}_{u},\tilde{m}_{c},\tilde{m}_{d},\tilde{m}_s )$ and only
one free parameter, namely, the CP violating phase $\Phi$.

We made a new $\chi ^{2}$ fit of the absolute value of the entries in
the mixing matrix, $|V_{CKM}^{th}|$, to the experimentally determined
values $|V_{CKM}^{exp}|$. We kept the mass ratios fixed at the values
given in (\ref{102}) and (\ref{104}). We varied only the CP violating
phase $\Phi$. The best  value of $\Phi$ was found to be $89.3^{\circ}$
corresponding to a minimum value of $\chi^{2}$ equal to 0.33. As explained at the end of section \ref{h5}, we may round off to 

\begin{eqnarray}\label{168}
\Phi^{*} = 90^{\circ}
\end{eqnarray}
witout spoiling the good quality of the fit.

The mixing matrix  ${{\bf V}^{th}_{CKM}}$, computed with this value of
$\Phi$ is

\begin{eqnarray}\label{170}
{\bf V}_{CKM}^{th}= \pmatrix{
0.9750+i0.0188 & -0.2020+i0.0906 & 0.0003+i0.0037 \cr
-0.0907+i0.2019 & 0.0188+i0.9742 & -0.0000+i0.0396 \cr
-0.0000-i0.0084 & -0.0000-i0.0388 & 0.0000+i0.9992 \cr
}.
\end{eqnarray}

The matrix of the moduli, computed from (\ref{170}), is

\begin{eqnarray}\label{172}
|{V^{th}}_{CKM}|=\pmatrix{
0.9752 & 0.2214 & 0.0037 \cr
0.2213 & 0.9744 & 0.0396 \cr
0.0084 & 0.0388 & 0.9992 \cr
},
\end{eqnarray}
which is to be compared with the experimental value\cite{24}
\begin{eqnarray}\label{174}
\mid {V^{exp}}_{CKM} \mid=\pmatrix{
0.9745-0.9760 & 0.217-0.224 & 0.0018-0.0045 \cr
0.217-0.224 & 0.9737-0.9753 & 0.036-0.042 \cr
0.004-0.013 & 0.035-0.042 & 09991-0.9994 \cr
}.
\end{eqnarray}
We see that the absolute values of the entries in the mixing matrix
computed from the theoretical expressions for ${\bf V}^{th}_{CKM}$, with
the values of the mass ratios given in (\ref{102}) and (\ref{104}) \cite{16} reproduce the central values of the experimentally determined entries in
$|{\bf V}^{exp}_{CKM}|$, almost exactly, well within the bounds of
experimental error. 

We also computed the Jarlskog invariant $J$ \cite{26}. The value obtained from
(\ref{100}) is
\begin{eqnarray}\label{176}
J^{th} =3.00 \times 10^{-5} ,
\end{eqnarray}
in good agreement with current data on CP violation in the
$K^{\circ}-\bar{K}^{\circ}$ mixing system \cite{24}.

The three inner angles of the unitarity triangle may now be readily
computed from the expressions ~(\ref{128})-(\ref{132}). We found the following values
\begin{eqnarray}\label{178}
\alpha = 84^{\circ}\qquad\qquad
\beta = 24^{\circ}\qquad\qquad
\gamma = 72^{\circ}
\end{eqnarray}

These three angles will be determined from $CP$ asymmetries in a
variety of weak $B$ decays at the forthcoming $B$ factories.

An estimation of the range of values of these angles compatible with
the experimental information on the absolute values of the matrix
elements of ${\bf V}^{exp}_{CKM}$, is given by S. Mele \cite{28} and
A. Ali \cite{29}. According to these authors, $79^{\circ}\leq \alpha
\leq 102^{\circ}$, $21^{\circ}\leq \beta \leq 28^{\circ}$ and 
$55^{\circ}\leq\gamma\leq 78^{\circ}$. We see that the value of
$\beta$ obtained in this work coincides almost exactly
with the central value of $\beta$ according to S. Mele \cite{28}, while our 
$\gamma$ is close to the upper limit given by S. Mele \cite{28} and $\alpha$ is in the allowed range given by these authors.

\section{Summary and conclusions}\label{h8}

In this work we derived theoretical expressions for the mixing matrix
${{\bf V}^{th}}_{CKM}$ from quark mass matrices ${\bf M}_{q}$ with a
modified Fritzsch texture. The mass matrices were built up adding
three terms ${\bf M}_{3q} ,{\bf M}_{2q}$ and ${\bf M}_{1q}$
corresponding to stages of less symmetry in a simple scheme for
breaking the flavour permutational symmetry.

The breaking pattern of the $S_{L}(3)\otimes S_{R}(3)$ symmetry down
to $S_{L}(2)\otimes S_{R}(2)$ was characterized in terms of the
parameter $Z^{1/2} = {(M_{2q,H})_{23}\over (M_{2q,H})_{22}}$   which is a
  measure of the amount of mixing of singlet and doublet irreducible
  representations of $S_{L}(3)\otimes S_{R}(3)$. This breaking pattern
  was classified in terms of the symmetric $(Z^{1/2}_{S}= 1/{\sqrt{8}})$,
  and antisymmetric $(Z^{1/2}_{A}=- \sqrt{8})$ representations of an
  auxiliarly group $\tilde{S}(2)$ of permutations of the two fields in
  the Yukawa term ${\bf \bar{q}}_{L,W}{\bf M}_{2q,W}{\bf q}_{R,W}$

A careful comparison of the theoretical expression for the absolute
values of the elements of the $CKM$ matrix with the experimentally
determined values of $|V_{CKM}^{exp}|$, $J^{exp}$ and the inner angles of the unitarity triangle $\alpha^{exp}$, $\beta^{exp}$ and $\gamma^{exp}$ gives a clear and precise indication on the existence of a preferred pattern for breaking the $S_{L}(3)\otimes S_{R}(3)$ flavour symmetry down to 
$S_{L}(2)\otimes S_{R}(2)$. The preferred or best symmetry breaking pattern is
characterized by

\begin{eqnarray}\label{184}
Z^{*1/2} =  \frac{1}{2}\left(Z^{1/2}_{S}-Z^{1/2}_{A}\right) =
\sqrt{\frac{81}{32}}.
\end{eqnarray}

Once the numerical value of $Z^{* 1/2}$ is fixed at
$\sqrt{\frac{81}{32}}$, the mass matrices ${\bf M}_{q}$ are functions of the
quark masses and only one phase. In consequence, the resulting best
theoretical $V_{CKM}$ matrix is parametrized in terms of the four
quark mass ratios
$(\tilde{m}_{u},\tilde{m}_{c},\tilde{m}_{d},\tilde{m}_{s})$ and only
one CP violating phase $\Phi$. The best value of $\Phi$ was found to
be 

\begin{eqnarray}\label{186}
\Phi = 90^{\circ} .
\end{eqnarray}
 The moduli of the matrix elements of the mixing matrix
 computed from the theoretical expression $V_{CMK}^{th}$ are in
 excellent agreement with all the experimentally determined absolute
 values of the $CKM$ matrix $|V_{CKM}^{exp}|$. For the Jarlskog
 invariant we found the value $J = 3.00\times 10^{-5}$ and for the inner angles of the unitarity triangle we found the values $\alpha= 84$, $\beta= 24$ and $\gamma= 72$ also in very good agreement with current data on CP violation in the $K^{\circ}-\bar{K}^{\circ}$ mixing system\cite{24} and the most recent data on oscillations in the $B^{\circ}_{s}-\bar{B}^{\circ}_{s}$ system \cite{28} and \cite{29}.
 
 In the standard electroweak model both the masses of the quarks as well as the weak mixing angles appear as free parameters. In this work, we have shown that, starting from the flavour permutational symmetry, a simple and explicit ansatz about the pattern of symmetry breaking leads to a parametrization of the
 $CKM$ mixing matrix in terms of four quark mass ratios $(m_{u}/m_{t},
 m_{c}/m_{t}, m_{d}/m_{b}, m_{s}/m_{b})$ and one CP violating phase in
 very good agreement with all the available experimental information on quark
 mixings and CP violation.

\section*{Acknowledgements}\label{h9}

We gratefully acknowledge many useful discussions with Prof. P.K. Kabir and Dr. M. Mondrag\'on.
One of us, E. R-J is indebted to Dr. J. R. Soto for help in the
numerical calculations. This work was partially supported by
DGAPA-UNAM under contract No. PAPIIT-IN110296 and by CONACYT,
(M\'exico) under contract No. 3909P-E9607.




\end{document}